\documentclass[aps,prapplied,longbibliography,twocolumn,amsmath,amssymb]{revtex4-1}

\usepackage{amsmath,amssymb,graphicx,xcolor,units,hyperref}

\newcommand{\dd}[1]{\mathrm{d}#1\,}

\renewcommand{\Re}{\mathop{\mathrm{Re}}}

\DeclareMathOperator{\csch}{csch}
\DeclareMathOperator{\arcosh}{arcosh}

\newcommand{\arccosh}{\mathop{\mathrm{arccosh}}}

\begin{document}

\title{Spectral characteristics of a fully-superconducting SQUIPT}

\author{P.~Virtanen}
\email{pauli.virtanen@nano.cnr.it}
\affiliation{NEST, Istituto Nanoscienze-CNR and Scuola Normale Superiore, I-56127 Pisa, Italy}

\author{A.~Ronzani}
\affiliation{NEST, Istituto Nanoscienze-CNR and Scuola Normale Superiore, I-56127 Pisa, Italy}

\author{F.~Giazotto}
\affiliation{NEST, Istituto Nanoscienze-CNR and Scuola Normale Superiore, I-56127 Pisa, Italy}

\date{\today}

\begin{abstract}
  We consider properties of a fully superconducting variant of the
  superconducting quantum interference proximity transistor, a magnetic
  flux sensor. We study the density of states in a finite-size
  superconducting metal wire in the diffusive limit, and how it
  depends on the phase gradient of the order parameter. We describe
  the dependence on the junction length and interface transparency,
  and discuss properties relevant for using the structure in magnetic
  flux detection applications.
\end{abstract}

\maketitle

\section{Introduction}

Superconductivity on the scale of the coherence length is sensitive to
its surroundings. This can be used to modulate the density of states
in a mesoscopic metal wire via magnetic flux, by imposing a phase
gradient via embedding the wire as a weak link in a superconducting
ring.  Detecting the modulation with a tunnel junction probe attached
to the weak link is the basis of the superconducting quantum
interference proximity transistor (SQUIPT), which can be used as a
magnetic field
sensor. \cite{giazotto2010-sqi,meschke2011-tsp,giazotto2011-hsq,strambini2014-pnw,giazotto2014-ppc,giazotto2013-qih,dambrosio2015-nmt,ronzani2014-hss}

Previous experiments mostly employed normal-state metal
wires. \cite{giazotto2010-sqi} For practical purposes, it however can
be advantageous if also the weak link is made of a superconducting
rather normal-state material: sample fabrication is simpler, quality
of the contacts can be better, and the quality of the energy gap can
improve, resulting to better sensitivity. The intrinsic
superconductivity however modifies the current-phase relation, which
determines what phase gradient can be imposed, and affects the
detailed form of the density of states.

The current-phase relation in SS'S junctions has been extensively
studied in the past, and is largely understood.
\cite{likharev79,golubov2004-cri,kupriyanov1981-ssp,kupriyanov1980,kupriyanov1982-iee}
The density of states (DOS) is moreover well-studied in the
normal-state case.
\cite{mcmillan1968-tms,ishii1972-tpj,golubov1996-qcb,zhou1998-dos,gueron96,sueur2008-pcs}
The DOS in a phase-biased superconducting wire, which is the case
relevant for the fully superconducting SQUIPT, appears to have
received somewhat less comment.
\cite{anthore2003-dos,gumann2007-mts,levyyeyati1995-sct}

\begin{figure}[t]
  \includegraphics{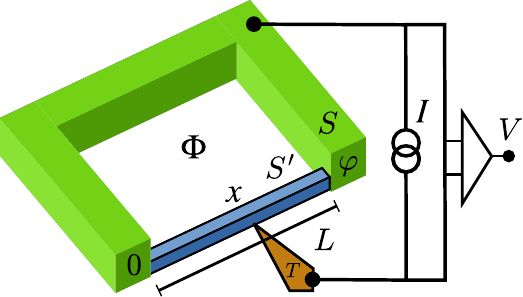}
  \caption{
    \label{fig:squipt}
    Superconducting quantum interference proximity transistor,
    consisting of a superconducting (S') weak link of length $L$
    embedded in a superconducting (S) SQUID ring. The magnetic flux
    $\Phi$ induces a difference of $\varphi$ in the phase of the
    order parameter across the junction, which in the weak link limit is
    $\varphi=2\pi\Phi/\Phi_0$.  The density of states in the weak link
    can be probed with a normal or superconducting tunnel probe (T).
  }
\end{figure}

In this work, we investigate the density of states and other
characteristics of a phase-biased superconducting wire, such as those
embedded as a part of a SQUIPT structure, as depicted in
Fig.~\ref{fig:squipt}.  We approach the problem of finding the
current-phase relation and spectral characteristics from a functional
minimization perspective. We describe the evolution of the
current-phase relation and the density of states between short and
long junction limits.  Finite-size effects in the DOS turn out to
decay as a power law with increasing system length, rather than
exponentially as they do for the current-phase relation.  We discuss
the effect of differing materials in the ring and wire parts, the
impact of interface transparency, and some practical consequences for
the S-SQUIPT application.

\section{Model}

We consider the SQUIPT setup depicted in Fig.~\ref{fig:squipt}, where a
mesoscopic conventional superconductor (S) metal wire of length $L$
and cross section ${\cal A}$ is embedded as a part of a
superconducting ring. The density of states in the wire is probed by
the current-voltage characteristics of a normal (N) or superconducting
tunnel junction.  The critical temperature of the ring material is
$T_{c,R}$, and that of the wire material is $T_{c,w}$, corresponding
to the superconducting coupling constants $\lambda_R$, $\lambda_{c}$
and the zero-temperature values $\Delta_{0,R}$, $\Delta_{0,w}$ of the
energy gap via the BCS relations.  Possible dependence of the
superconducting coupling on the film thickness is included in these
parameters.  We assume the dimensions of the wire are small compared
to the ring, so that the presence of the wire has small effect on
the superconductivity of the ring.

The superconducting properties in the diffusive limit at equilibrium
are conveniently described by the nonlinear $\sigma$-model.
\cite{belitz1994-amt,altland1998-ftm,taras-semchuk2001-qif,yurkevich2001-nsm}
The approach encompasses the well-known quasiclassical Green function
theory \cite{belzig1999-qgf} of diffusive superconducting systems as a
special case, but can also be used to study eg. fluctuation effects.
Here, we consider only equilibrium properties of quasi-1D systems in
the semiclassical approximation --- in this case the advantage is in
directly specifying the problem in variational form. At the classical
saddle points the free energy contribution is $\delta{}F={\cal
  A}\nu_F[F_b + \int_0^L\dd{x}{\cal F}_0]$, where the density reads
(cf. Ref.~\onlinecite{eilenberger1968-tog})
\begin{align}
  {\cal F}_0
  &=
  \frac{
    |\Delta|^2
  }{\lambda{}}
  + 
  2\pi{}T
  \sum_{\omega_n>0}
  \{
  \frac{D(\partial_x\chi-2A)^2}{2}\sin^2\tilde{\theta}
  +
  \frac{D(\partial_x\tilde{\theta})^2}{2}
  \\\qquad&\notag
  +
  2\omega_n(1 - \cos\tilde{\theta})
  -
  2|\Delta|\cos(\chi-\phi)\sin\tilde{\theta}
  \}
  \,.
\end{align}
Here, $\omega_n=2\pi T(n+\frac{1}{2})$ are Matsubara frequencies,
$\nu_F$ the DOS per spin at the Fermi level, and $A$ the vector potential.  $D$
is the diffusion constant and $\Delta=|\Delta|e^{i\phi}$ the order
parameter.  Moreover, $\lambda$ is the superconducting interaction
constant.  The parameters $\chi$ and $\tilde{\theta}$ are real-valued,
and at saddle points related to the quasiclassical anomalous Green
function
$f(x,\omega_n)=e^{i\chi(x,\omega_n)}\sin\tilde{\theta}(x,\omega_n)$. \cite{belzig1999-qgf}
Here and below, we use units $k_B=e=\hbar=1$.
For taking the weak
coupling limit $\lambda\to0$, $\Delta_{w,0}=\text{const.}$,
it is useful to add and subtract terms to eliminate the implicit cutoffs,
\begin{align}
  \label{eq:replica-action}
  {\cal F}_0
  &\simeq
  |\Delta|^2\ln\frac{T}{T_c}
  +
  2\pi{}T
  \sum_{\omega_n>0}\{
  \frac{|\Delta|^2}{\omega_n}
  +
  \frac{D(\partial_x\chi-2A)^2}{2}\sin^2\tilde{\theta}
  \\\qquad&\notag
  +
  \frac{D(\partial_x\tilde{\theta})^2}{2}
  +
  2\omega_n(1 - \cos\tilde{\theta})
  -
  2|\Delta|\cos(\chi-\phi)\sin\tilde{\theta}
  \}
  \,.
\end{align}
The dependence on $\lambda$ is now contained in $T_c$, assumed to
be of the BCS weak-coupling form.

The connection of the wire to the superconducting ring is described
via a tunneling boundary term, \cite{kupriyanov1988-iob,altland2000-ftm}
\begin{align}
  \label{eq:bc-action}
  F_b 
  &=
  \frac{2\pi{}TD}{r}\sum_{\omega_n>0}\sum_{j=\pm}[1-\cos\tilde{\theta}(x_j)\cos\tilde{\theta}_{Sj}
  \\\notag&
  -\cos(\chi(x_j)-\chi_{Sj})\sin\tilde{\theta}(x_j)\sin\tilde{\theta}_{Sj}]
  \,,
\end{align}
where $r=2R_I\mathcal{A}D\nu_F$ is the ratio of interface resistance $R_I$
to the resistance per length of the wire, and $\tilde{\theta}_{S\mp}=\arctan(\Delta_R/\omega_n)$,
$\chi_{S-}=0$ and $\chi_{S+}=\varphi$ are the values inside the ring
at the left ($x_-=0$) and right ($x_+=L$) interfaces.

Requiring variation of Eq.~\eqref{eq:replica-action}
vs. $\tilde{\theta}$ and $\chi$ to vanish, and analytical continuation
to real axis $i\omega\mapsto{}E+i0^+$ and defining
$\theta\equiv{}-i\tilde{\theta}$ produces the standard quasiclassical
real-time description of the system via the Usadel equation,
\cite{usadel1970-gde,belzig1999-qgf} which can be written in the form
\begin{gather}
  \label{eq:usadel}
  D\partial_x^2\theta
  =
  -2iE\sinh\theta + \frac{D(\partial_x\chi)^2}{2} \sinh2\theta
  \\\notag\qquad
  + 2i|\Delta|\cos(\phi-\chi)\cosh\theta
  \,,
  \\
  D\partial_x\cdot(\partial_x\chi \sinh^2\theta) = -2i|\Delta|\sin(\chi-\phi)\sinh\theta
  \,.
\end{gather}
The self-consistency equation for the order parameter $\Delta$ 
is obtained similarly,
\begin{align}
  \label{eq:selfcons}
  |\Delta| \ln\frac{T}{T_c}
  = 
  2\pi i T\sum_{\omega_n>0}[e^{i(\chi-\phi)}\sinh\theta - \frac{|\Delta|}{E}]\rvert_{E=i\omega_n}
  \,.
\end{align}
The boundary term generates the boundary
conditions, \cite{kupriyanov1988-iob}
\begin{align}
  \label{eq:kl}
  \mp{}r\partial_x\chi
  &=
  \sin(\chi-\chi_{S\mp})\frac{\sinh\theta_{S\mp}}{\sinh\theta}
  \,,
  \\
  \mp{}r\partial_x\theta
  &=
  \sinh\theta\cosh\theta_{S\mp}
  -\cos(\chi-\chi_{S\mp})\cosh\theta\sinh\theta_{S\mp}
  \,,
\end{align}
at the left ($-$) and right ($+$) interfaces.  In the clean-interface
limit $r\to0$ these reduce to continuity conditions
$\chi(x_\pm)=\chi_{S\pm}$,
$\theta(x_\pm)=\theta_{S\pm}$. Note also that in this case
$F_b\to0$ as $r\to0$.

The reduced density of states
$N(E,x)=\nu(E,x)/\nu_F=\Re\cosh\theta(E,x)$ and current
$I(x)=-\frac{\delta{}F}{\delta{}A}\rvert_{A=0}=4\pi{}T\mathcal{A}D\nu_F\sum_{\omega_n>0}\partial_x\chi\sin^2\tilde{\theta}$
follow directly from the solutions of the equation set.

It is well known that there are multiple classical solutions,
corresponding to different windings of the superconducting phase along
the superconducting wire.  In the numerical solution of the equation
set, to handle this and to obtain also the solutions along the
unstable branches, we use the pseudo-arclength continuation method
applied on the self-consistency equation. This method is useful for
tracing a continuous solution branch $(\varphi, \Delta)$ without
requiring the existence of a single-valued function $\Delta(\varphi)$
(see Appendix~\ref{sec:arclength} for details).

Within the approximations here, the configuration minimizing
$\delta{}F$ should be considered as the stable solution. Along a
continuous solution branch, this can also be evaluated via a standard
relation,
\begin{align}
  \label{eq:cpr-energy}
  \delta{}F[X_*(\varphi);\varphi]-\delta{}F[X_*(0);0]
  =
  \frac{1}{2}
  \int^{\varphi}_0\dd{\varphi'}I[X_*(\varphi');\varphi']
\end{align}
based on the current evaluated at a stationary point
$X_*=(\tilde{\theta}_*,\chi_*,|\Delta_*|,\phi_*)$. This follows from
Eqs.~\eqref{eq:replica-action},\eqref{eq:bc-action} by noting the
gauge transform
$F[\tilde{\theta},\chi,|\Delta|,\phi;\varphi,A]=F[\tilde{\theta},\chi-\xi,|\Delta|,\phi-\xi;\varphi-\xi(L),0]$
for $\xi(x)=2\int_0^x\dd{x'}A(x')$ and that
$\delta{}F/\delta{}X\rvert_{X=X_*}=0$.  Energy differences between
disconnected branches however need to be determined from
Eqs.~\eqref{eq:replica-action},\eqref{eq:bc-action}.

\section{Density of states and current}

\begin{figure}
  \includegraphics{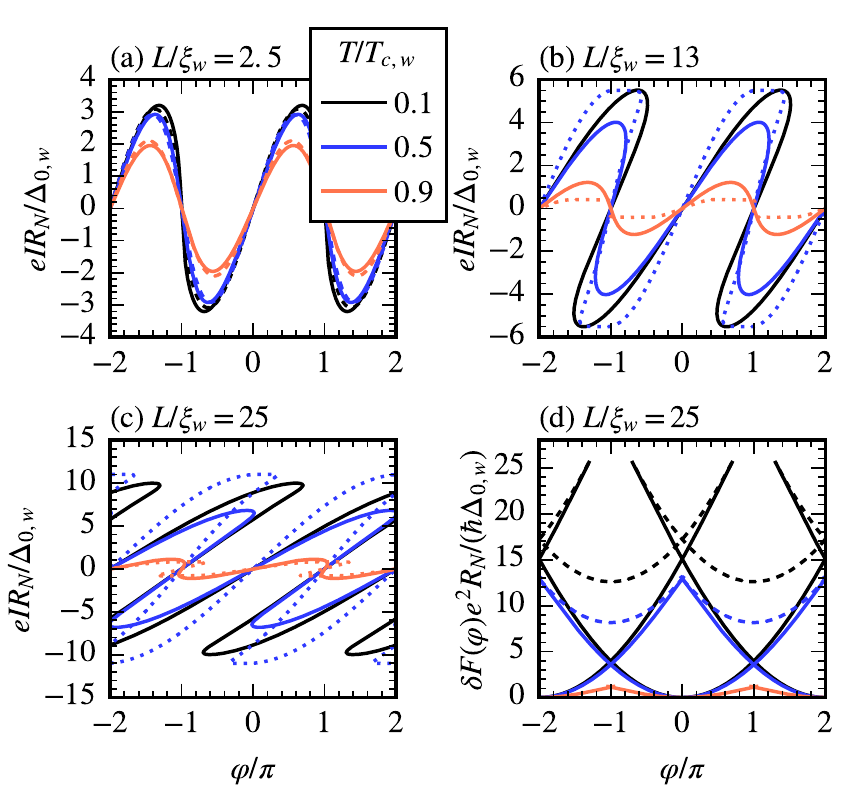}
  \caption{
    \label{fig:cpr}
    (a-c)
    Current-phase relation vs. junction length and
    temperature ($T/T_{c,w}=0.1$, $0.5$, $0.9$), for $T_{c,R}/T_{c,w}=1.5$.
    Here, $\xi_{w}=\sqrt{D/(2\pi{}T_{c,w})}$.
    Analytical short-junction result \cite{kulik1975-cmt,likharev79} (dashed) and 
    the GL result discussed in the text (dotted) are also shown.
    (d)
    Free energy change corresponding to curves in (c),
    as obtained from Eq.~\eqref{eq:replica-action} or~\eqref{eq:cpr-energy}.
    Backward branches are shown with dashed lines.
  }
\end{figure}

The density of states in S-SQUIPT is sensitive to the phase point
where the system is biased. To specify this, we need to comment on the
supercurrent in the junction.  The form of the current-phase relation
(CPR) in superconducting strips is well studied under the
approximations outlined
above. \cite{likharev79,golubov2004-cri,kupriyanov1981-ssp,kupriyanov1980,kupriyanov1982-iee}
CPRs computed from the Usadel equation are illustrated in
Fig.~\ref{fig:cpr}, for reference; see also
Ref.~\onlinecite{kupriyanov1981-ssp}.
\footnote{
  For the parameters of Ref.~\onlinecite{kupriyanov1981-ssp}, we obtain
  coinciding results for the CPRs.
}
For junctions short compared to the
coherence length, $L\ll{}\xi_w\equiv\sqrt{D/(2\pi{}T_{c,w})}$,
the CPR is a (deformed) sinusoid, described by a known analytical solution.
\cite{kulik1975-cmt,likharev79} In long junctions, the CPR becomes multivalued,
corresponding to multiple winding of the order parameter phase.
At temperatures close to $T_c$, the form can be found from Ginzburg-Landau equations. \cite{likharev79,langer1967-irt}
For $L\gg{}\xi_{GL}=\sqrt{D\pi/[8(T_c-T)]}$ one expects a CPR
$I=(\mathcal{A}\sigma_N\pi\Delta^2/(4TL))\varphi[1 - (\varphi\xi_{GL}/L)^2]$, up
to the point $\varphi<\varphi_{\rm max}\approx{}L/(\sqrt{3}\xi_{GL})$. After this, the
solution transitions to a backward branch reaching to $\varphi=\pi$,
$I=0$,
\begin{align}
  I &= \frac{\mathcal{A}\sigma_N\pi\Delta^2}{2T\xi_{GL}} a\sqrt{a-k}
  \,,
  \;
  \varphi =
  \frac{L}{\xi_{GL}}\sqrt{a-k}
  + \frac{1}{a}\arcsin\sqrt{\frac{k}{a}}
  \,,
\end{align}
for $k\in[0,1/2]$, $a=(k+1)/3$. The phase gradient becomes more non-uniform,
corresponding to the formation of a phase slip center in the middle of
the junction. \cite{langer1967-irt,ivlev1984-tcs}

\begin{figure}
  \includegraphics{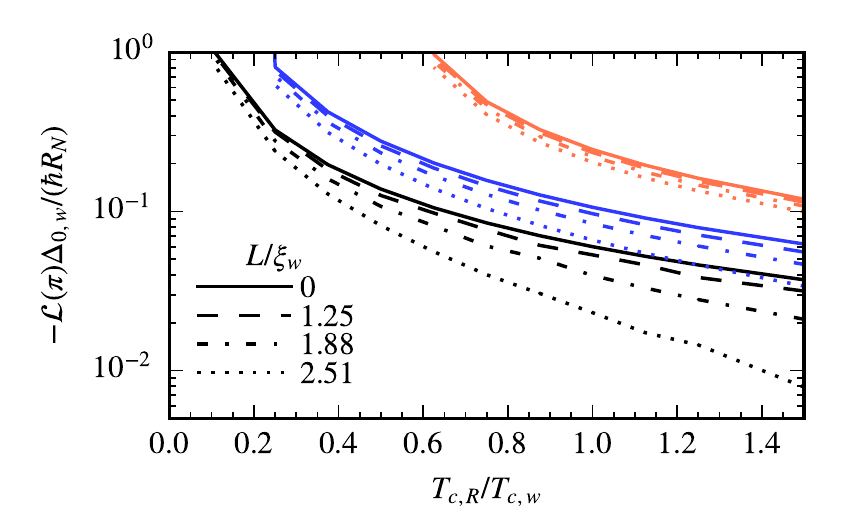}
  \caption{
    \label{fig:pi-slope}
    Normalized inductance of the superconducting wire at $\varphi=\pi$,
    for different $T_{c,R}$ and $L$, at $T/T_{c,w}=0.01$, $0.1$, $0.5$
    (from bottom to top). Solid lines denote
    the short-junction limit~\eqref{eq:inductance-an}.
  }
\end{figure}

In long junctions the CPR allows several saddle point solutions for
the current $I$ when $\varphi$ is fixed.  The backward solution branches
are unstable and not directly accessible. Which of the remaining
multiple solutions are accessible depends on the rate $\Gamma$ of
phase slips by which the system can transition to lower-energy
states. \cite{langer1967-irt,arutyunov2008-sio} The free energy
barrier $U$ in the thermal activation rate $\Gamma\propto{}e^{-U/T}$
can be estimated from the difference between the unstable and
metastable branch energies in Fig.~\ref{fig:cpr}d and is of order
$U\sim{}\hbar{}\Delta_{0,w}/(e^2R_\xi)$, $R_\xi\equiv\xi_w{}R_N/L$.
\cite{langer1967-irt,ivlev1984-tcs,arutyunov2008-sio,zharov2007-mtt,semenov2010-mtp}
Quantum phase slip rate in nanowires is $\Gamma\propto{}e^{-a\hbar{}/(e^2R_\xi)}$
with $a\sim1$.  \cite{arutyunov2008-sio} For typical S-SQUIPT
cross-section $\unit[100]{nm}\times\unit[30]{nm}$ and
high-conductivity material, $R_\xi\sim\unit[1]{\Omega}$, so that both
rates are effectively zero except close to $\varphi\approx\varphi_{\rm
  max}$ where the CPR bends backward, or at high temperatures. For low
relaxation rates, the experimental CPR has magnetic hysteresis, and
the different branches can be accessed by sweeping the magnetic flux.

In short junctions where the CPR is single-valued, the kinetic
inductance of the wire decreases around $\varphi=\pi$, and can
become small compared to that of the ring, inducing behavior similar
to hysteretic rf-SQUIDs. \cite{meschke2011-tsp} To quantify this, we show in
Fig.~\ref{fig:pi-slope} the Josephson inductance ${\cal
  L}=\hbar/(2e\partial_\varphi{}I)$ at $\varphi=\pi$ for different
temperatures and wire lengths.  In the short-junction limit
$L\ll{}\xi_w$, from the known expression for the CPR, \cite{kulik1975-cmt}
after some rewriting we have
\begin{align}
  \label{eq:inductance-an}
  \frac{1}{
    {\cal L}(\pi)
  }
  =
  -
  \frac{
    \pi\Delta_R
  }{
    \hbar{}R_N
  }
  \int_0^{\Delta_R/(2T)}
  \dd{z}\frac{\tanh z}{z}
  \,.
\end{align}
For $T\lesssim{}\Delta_R/2$, we have
${\cal{}L}(\pi)^{-1}\simeq{}\frac{\pi\Delta_R}{\hbar{}R_N}\ln\frac{T}{2T_{c,R}}$.
As visible in Fig.~\ref{fig:pi-slope}, for $L>\xi_{w}$ the normalized inductance
in general further decreases from the short-junction value, regardless
of the ratio of $T_{c,R}$ and $T_{c,w}$. However, the value is tunable
by material choices.


Let us now consider the density of states (DOS) in the junction.
Analytical solutions to Eq.~\eqref{eq:usadel} providing access to the
DOS are known in the long-junction ($L\gg{}\xi$) and
short-junction ($L\ll{}\xi$) limits. In the short-junction
limit, we have a well-known result
(see eg. Ref.~\onlinecite{heikkila2002-sdo}),
%
\begin{align}
  \label{eq:short}
  N(x,E)
  &=
  \Re
  \sqrt{\frac{E^2}{E^2-\Delta_R^2\cos^2\frac{\varphi}{2}}}
  \cosh
  \biggl(
  \frac{2x-L}{L}
  \\\notag&\qquad
  \times\arcosh\sqrt{\frac{E^2-\Delta_R^2\cos^2\frac{\varphi}{2}}{E^2-\Delta_R^2}}
  \biggr)
  \,,
\end{align}
\emph{independent} of the superconductivity of the wire itself.  For long
wires ($L\gg\xi_{w}$), on the other hand, the result converges to the density of
states of a bulk superconductor, affected by depairing from the
supercurrent flow. \cite{anthore2003-dos,romijn1982-cpb,kupriyanov1980,kupriyanov1982-iee}
This can be described as a depairing
rate, \cite{abrikosov1960} $g^2=D(\partial_x\phi)^2$,
given a constant phase gradient $\partial_x\phi$ small enough:
\begin{align}
  \label{eq:bulkg}
  N(E)=\Re\cosh\theta_0
  \,,
  \quad
  |\Delta|\coth\theta_0
  =
  E + i\frac{g^2}{2}\cosh\theta_0
  \,.
\end{align}
For long junctions, the parameter can be estimated from
GL solutions \cite{langer1967-irt} to be
$g^2\approx{}D(\varphi/L)^2$ along the forward
branch, and
$g^2\lesssim{}D\max(\partial_x\phi)^2\approx{}D[1+(L/\xi - 2\sqrt{2})^2/(\varphi-\pi)^2]/4$
on the backward branch close to $\varphi=\pi$.  The approximation in
Eq.~\eqref{eq:bulkg} is however not expected to work as well along the
backward branch, as the phase gradient is not uniform due to the
formation of the phase slip center.

\begin{figure}
  \includegraphics{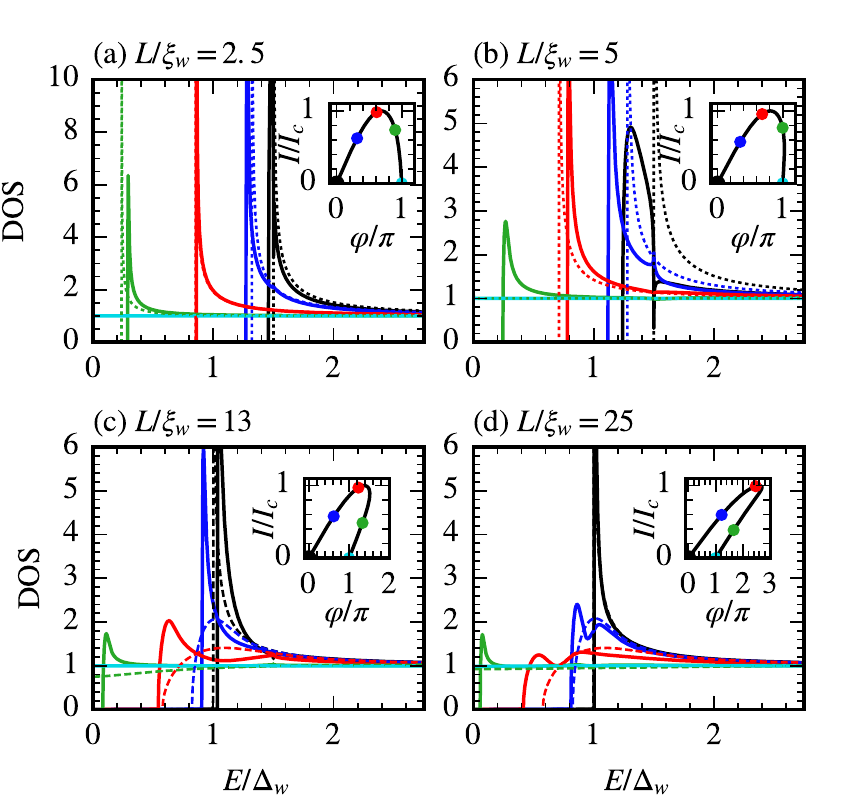}
  \caption{
    \label{fig:dos}
    Density of states in an SS'S junction at $x=L/2$, for different phase
    differences and junction lengths $L$. Here, $T_{c,R}=1.5T_{c,w}$,
    and $T/T_{c,w}=0.1$.
    Dotted lines in (a) indicate Eq.~\eqref{eq:short}, and dashed in (b),(c),(d)
    Eq.~\eqref{eq:bulkg}.
    Insets: current-phase relation and the points corresponding to the
    densities of states shown.
  }
\end{figure}

\begin{figure}
  \includegraphics{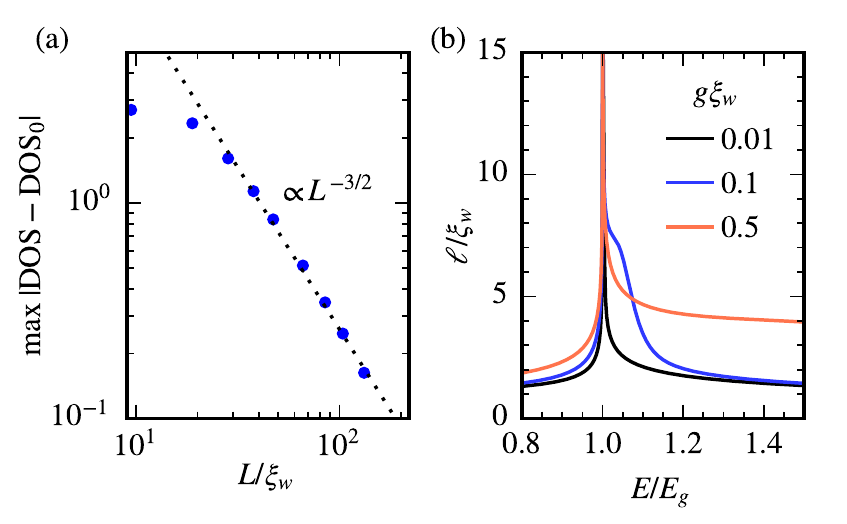}
  \caption{
    \label{fig:ddos-scaling}
    (a)
    Maximum difference in density of states between
    Eq.~\eqref{eq:bulkg} and the solution in SS'S configuration,
    assuming $\Delta(x)=|\Delta_0|e^{igx}$ with $g=0.1$.  
    (b)
    Perturbation decay length $\ell$.
  }
\end{figure}

The cross-over from the short-wire to the long-wire limit is
illustrated in Fig.~\ref{fig:dos}. We can observe that the
short-junction solution is fairly accurate up to $L\sim\xi_{w}$,
except in narrow energy regions around the gap edges
$E=E_g(\varphi,L)$.
As the length increases, the DOS converges towards that of a bulk
superconductor affected by depairing from the superflow (dashed lines).
However, the peaks at the DOS gap edge vanish only
rather slowly as $L/\xi_{w}\to\infty$ and are not present in the
long wire limit described by Eq.~\eqref{eq:bulkg}. The cross-over
is not very rapid, indeed, the rate of the decay is a power law rather
than exponential, as shown in Fig.~\ref{fig:ddos-scaling}a.

The physical mechanism giving rise to the finite-size effects in the
DOS are the Andreev reflections at the ring interfaces, $x=0$, $x=L$.
In a normal metal wire, this effect decoheres on a certain decay
length scale $\ell(E)\propto{}\sqrt{\hbar{}D/E}$, which can be
understood to originate from energy dependent phase shifts between
electrons and the almost retroreflected holes
(cf. e.g. Ref.~\onlinecite{pannetier2000-ara} for review). The
possibility of Andreev reflections also inside the superconducting
wire itself however alters $\ell(E)$, e.g., inhibiting low-energy
electrons from reaching the ring boundary. Moreover, in a
superconductor the decay length diverges around the gap edge $E_g$,
\cite{larkin1972-dos} instead of around $E=0$. This can be seen by
considering small perturbations $\theta=\theta_0+\eta$,
$\chi=\phi+\alpha$, $|\eta|\ll1$, $|\alpha|\ll1$ around the uniform
solution $(\theta_0,\phi)$ described by Eq.~\eqref{eq:bulkg}.
Substituting such ansatz in Eq.~\eqref{eq:usadel} and linearizing
around $(\theta_0,\chi_0)$, produces solutions of the form
$\eta,\chi\propto{}e^{\pm{}x/\ell{}}$, where $\ell$ are the decay
lengths. Consequently, the factor $\sim{}e^{-L/(2\ell)}$ indicates how
much the boundary conditions affect the solution at the center of the
wire. The linearized equation can be written here as
$\partial_x(\partial_x\eta,\eta,\partial_x\alpha,\alpha)^T=M(\partial_x\eta,\eta,\partial_x\alpha,\alpha)^T$
where the matrix $M$ is
\begin{align}
  M = 
  \begin{pmatrix}
    0 & \ell_0^{-2} & g\sinh2\theta_0 & 0 \\
    1 & 0 & 0 & 0 \\
    -2g\coth\theta_0 & 0 & 0 & -2i|\Delta|\csch\theta_0 \\
    0 & 0 & 1 & 0
  \end{pmatrix}
  \,,
\end{align}
where $\ell_0^{-2}=-2i|\Delta|\csch\theta_0+g^2\sinh^2\theta_0$.  The longest
decay length is given by the eigenvalue with the smallest real part,
$\ell^{-1}=\min{}|\Re\lambda|$. They are solutions to
\begin{align}
  \lambda^{4} + (5g^2\cosh^2\theta_0-g^2-2\ell_0^{-2})\lambda^{2} = \frac{2i|\Delta|\ell_0^{-2}}{\sinh\theta_0}
  \,.
\end{align}
The energy dependence of $\ell(E)$ is shown in
Fig.~\ref{fig:ddos-scaling}b, with a divergence clearly visible.  From
Eq.~\eqref{eq:bulkg} it follows that the gap edge is located at
\cite{abrikosov1960} $E_{g}=\Delta[1 - (g^2/2\Delta)^{2/3}]^{3/2}$,
and $\theta_{0}(E_{g})=\frac{-i\pi}{2} + \arccosh[(2\Delta
  g^{-2})^{1/3}]$.  Consequently, $\ell_0^{-2}(E_{g}) = 0$, so that
also $\ell^{-2}(E_{g})=0$. Around the gap edge, the length scale
diverges as $\ell\propto(E-E_{g})^{-1/4}$. As the decay length is
larger than $L$ in an energy range of width
$\delta{}E\propto{}L^{-4}$, the deviations from Eq.~\eqref{eq:bulkg}
at the middle of a long wire do not decay exponentially with
increasing $L$, and the scaling of Fig.~\ref{fig:ddos-scaling}a can
occur.


%

\begin{figure}
  \includegraphics[width=\columnwidth]{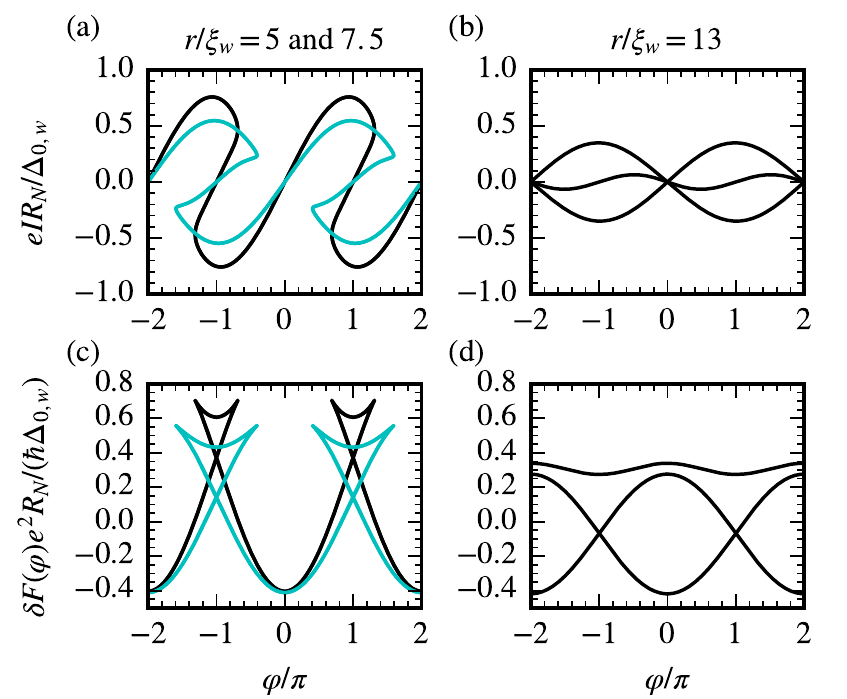}
  \caption{
    \label{fig:interface-transparency-cpr}
    (a),(b)
    CPR in SIS'IS system for $L/\xi_{w}=2.5$, $T/T_{c,w}=0.1$, and $T_{c,R}/T_{c,w}=1.5$,
    The dimensionless interface resistances are $r/\xi_w=5$ (black), $7.5$ (cyan) on the left panel,
    and $13$ on the right.
    Higher energy solutions may also exist, but are not shown.
    (c),(d)
    Corresponding free energy contribution from Eqs.~(\ref{eq:replica-action}--\ref{eq:bc-action}).
  }
\end{figure}

\subsection{Interface resistance}

Imperfect interface transparency in SQUIPT influences both the CPR and
the DOS. In particular, it enables additional stationary solutions,
where phase drops across the barriers at the interfaces.  With
increasing interface resistance $r$, the current-phase relation
crosses over to $I(\varphi)\simeq{}\pm{}I_c\sin(\varphi/2)$, that of
two Josephson junctions in series. \cite{kupriyanov1988-iob,levyyeyati1995-sct} For two
identical Josephson junctions and the superconducting wire in series,
with phase drop $\phi_1$ at the interfaces, the free energy is
\begin{align}
  \label{eq:toy-f}
  \delta{}F(\varphi,\phi_1) \sim -2E_{J}(r)\cos(\phi_1) - E_{J,\rm wire}\cos(\varphi-2\phi_1)
  \,.
\end{align}
For $r\to0$ ($E_J(r)\gg{}E_{J,\rm wire}$), lowest-energy solutions
have $\phi_1\approx2\pi{}n$ and $I(\varphi)\approx{}I_{\rm wire}(\varphi)$,
whereas for $r\to\infty$ ($E_J(r)\ll{}E_{J,\rm wire}$),
$\phi_1\approx{}\varphi/2+\pi{}n$ and consequently
$I(\varphi)\approx{}\pm{}I_J\sin(\varphi/2)$.

The cross-over is illustrated in
Fig.~\ref{fig:interface-transparency-cpr}(a),(b).  A nonzero but small
$r$ only effectively adds to the length of the junction, as shown in
panel (a) compared to Fig.~\ref{fig:cpr}a. A larger $r$ splits the
solution to disconnected branches as seen in panel (b) where there are
three different solutions at $\varphi=0$ yielding $I=0$. These
correspond to: (i) no phase drop across the wire or interfaces, (ii)
$\pi$ phase drop at both interfaces, and (iii) similar to the second
solution but smaller $|\Delta|$.  As $\varphi$ is varied continuously
from $0$ to $2\pi$, the solution (i) transforms to (ii) and vice
versa. The solution (iii) is not connected to the two; instead, when
$\varphi$ varies from $0$ to $\pi$, a phase slip center forms in the
center of the wire; cf. also Ref.~\onlinecite{levyyeyati1995-sct}.
The evolution is somewhat more clear in the free energy shown in
panels (c),(d). The energy differences between the solutions in (d)
must be obtained from
Eqs.~(\ref{eq:replica-action}--\ref{eq:bc-action}) instead of
Eq.~\eqref{eq:cpr-energy} as not all are continuously connected.  Note
that little qualitative change occurs in the lowest-energy
solution.

\begin{figure}
  \includegraphics{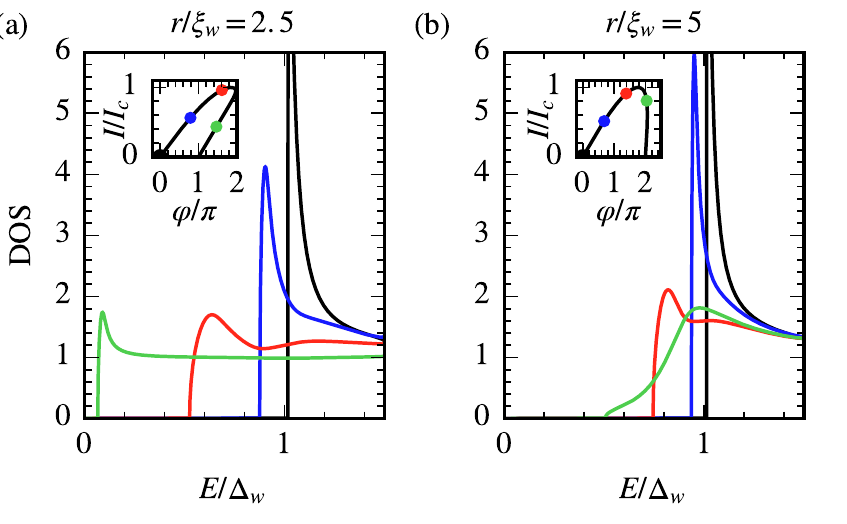}
  \caption{
    \label{fig:interface-transparency}
    DOS in SIS'IS system, for $L/\xi_w=13$; other parameters
    are as in Fig.~\ref{fig:dos}.
    Position along the solution branch is shown in the inset.
  }
\end{figure}

\begin{figure}
  \includegraphics{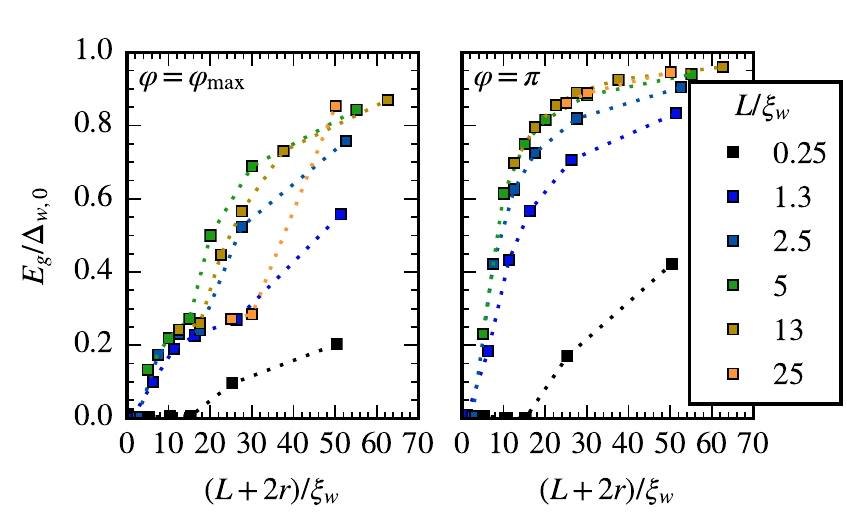}
  \caption{
    \label{fig:gap-modulation}
    Energy gap $E_g$ achieved at $\varphi=\varphi_{\rm max}$
    where CPR starts to bend back, and at $\varphi=\pi$
    in the lowest-energy state.
    Wire lengths and interface resistances are varied,
    other parameters are as in Fig.~\ref{fig:dos}.
  }
\end{figure}

The proximity effect from the ring diminishes with increasing barrier
height $r$, and controlling the properties of the weak link via the
superconducting ring becomes less effective. This is illustrated in
Fig.~\ref{fig:interface-transparency} which shows the modulation of
the density of states for a long junction with two different interface
resistances.  As $r$ increases, and the critical current of the
interfaces becomes small compared to that of the wire, the DOS
approaches that of the bulk superconductor.

For the S-SQUIPT device application, a relevant metric is how much the
modulation of the superconducting gap is suppressed.  In
Fig.~\ref{fig:gap-modulation}, we show the smallest magnitude of the
energy gap achievable with phase biasing. Based on the above
discussion, the minimum is achieved at the point
$\varphi=\varphi_{\rm{}max}$ where the CPR transitions to the backward
branch. In the lowest-energy state, the minimum gap is achieved at
$\varphi=\pi$, which is also shown.  We can note that in
Fig.~\ref{fig:gap-modulation} the curves for different wire lengths
$L$ tend to collapse onto a single curve for $L\gtrsim{}2.5\xi_w$. The
interface resistance in this case acts similarly as an extension of
the junction length by $2r$, consistent with the increase in the total
resistance, although at $\varphi=\varphi_{\rm{}max}$ the behavior is
complicated by the crossover illustrated in
Fig.~\ref{fig:interface-transparency-cpr}.

%

\section{Device performance}

\begin{figure}
  \includegraphics{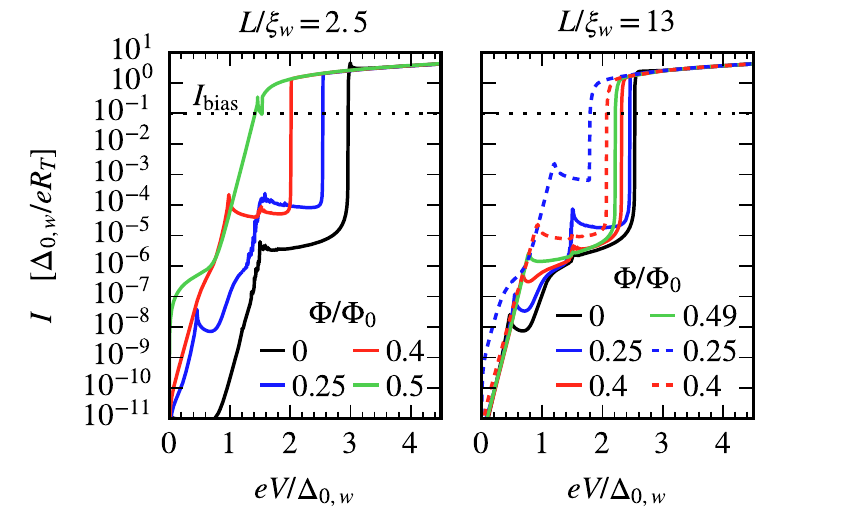}
  \caption{
    \label{fig:iv}
    Predicted IV characteristics of devices with $L=2.5\xi_w$ and $L=13\xi_w$ junctions,
    at $T/T_{c,w}=0.1$ and $T_{c,R}=1.5T_{c,w}$. For $L=13\xi_w$, the IV characteristics
    has flux hysteresis for $|\Phi-n\Phi_0|\gtrsim0.25\Phi_0$.
  }
\end{figure}

We can now discuss the performance of the SS'S devices in a
magnetometric measurement.  In this mode, the tunnel contact [(T) in
  Fig.~\ref{fig:squipt}] connected to the S' wire is current biased to
a working point $I=I_{\rm bias}$. The tunneling current depends on the DOS,
\begin{align}
  I(V,\Phi)
  &=
  \frac{1}{R_T}
  \int_{-\infty}^\infty
  \dd{E}N(E,\Phi)N_{\rm probe}(E-V)\times
  \\\notag&\qquad\times
  [f(E) - f(E-V)]
  \,,
\end{align}
and the variation of the observed voltage $V(I_{\rm bias},\Phi)$ as a function
of the flux $\Phi$ can be used to infer $\Phi$.  The sensitivity can
be characterized by the flux-voltage transfer function
\begin{align}
  {\cal F}(\Phi) = \frac{\dd{V}}{\dd{\Phi}}
  \,.
\end{align}
The resolution is intrinsically limited by the voltage noise in the
probe junction, which can be described by an equivalent flux noise
\begin{align}
  S_{\Phi,T}
  =
  \frac{S_V}{{\cal F}(\Phi)^2}
  \,,
\end{align}
where $S_V=(\frac{\dd{V}}{\dd{I}})^2S_I$ and
$S_I=2eI\coth\frac{eV}{2k_BT}$ is the tunneling shot
noise. Preamplifier voltage noise will also give a similar
contribution.

\begin{figure}
  \includegraphics{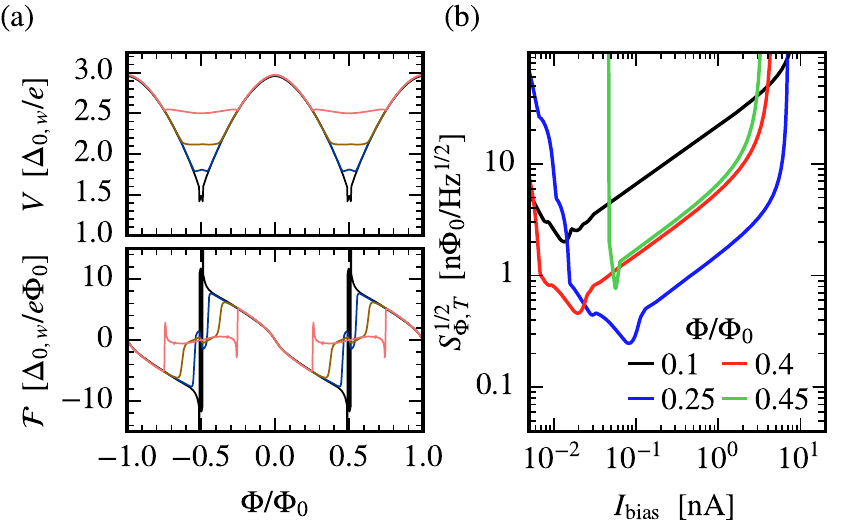}
  \caption{
    \label{fig:F}
    (a)
    $V$ vs. $\Phi$ corresponding to Fig.~\ref{fig:iv}
    and $L=2.5\xi_w$,
    for $eR_TI_{\rm bias}/\Delta_{0,w}=2$, $1.5$, $1$, $0.2$ (from top to bottom),
    and the corresponding transfer function ${\cal F}$.
    (b)
    Equivalent flux noise $\sqrt{S_{\Phi,T}}$ from tunnel junction, for
    $\Delta_{0,w}=\unit[200]{\mu{}eV}$ and $R_T=\unit[100]{k\Omega}$.
  }
\end{figure}

The current-voltage characteristic $I(V,\Phi)$ is shown in
Fig.~\ref{fig:iv} for the short $L=2.5\xi_w$ and long $L=13\xi_w$
junction cases.  Here, we assume the tunnel probe is superconducting,
with $T_{c,\mathrm{probe}}=T_{c,R}$.  For the short device fixed at
$I=I_{\rm bias}$, the external flux corresponds to a single measured
voltage value in a wide bias range, shown in Fig.~\ref{fig:F}a.  The
long-junction device on the other hand has magnetic hysteresis in the
range $|\Phi-n\Phi_0|\gtrsim0.25\Phi_0$, see Fig.~\ref{fig:cpr}, where
each flux value is associated with two possible $V$. Which one is
realized depends on the initialization of the device. Note that this
assumes the relaxation rates $\Gamma(\Phi)$ of metastable states are
very low, as was estimated in Sec.~III. If $\Gamma(\Phi)$ is not low
compared to measurement time scales, transitions can contribute
additional voltage noise, which reduces the usefulness of the
device. We will not consider this case here.

The voltage response at different bias currents and the corresponding
transfer function are shown in Fig.~\ref{fig:F}a for a short-junction
device. The overall behavior and magnitude of ${\cal F}$ is relatively
similar to an N-SQUIPT in this case. \cite{giazotto2011-hsq} For
long-junction devices, as also visible in Fig.~\ref{fig:iv}, we expect
decreasing device sensitivity with increasing system length, as the gap
suppression depends on the phase gradient which scales $\propto{}1/L$.

Calculated equivalent flux noise from the tunnel junction is displayed
in Fig.~\ref{fig:F}b, where we chose representative values for $R_T$ and
$\Delta_{0,w}$.  The tunnel junction noise is of the order of
$S_{\Phi,T}^{1/2}\sim\unit[10]{n\Phi_0/\sqrt{Hz}}$ at bias currents of
order $\unit[1]{nA}$.  Preamplifier voltage noise for a typical value
of $S_{V,\mathrm{pre}}^{1/2}\approx\unit[1]{nV/\sqrt{Hz}}$ on the other hand yields
$S_{\Phi,\mathrm{pre}}^{1/2}\gtrsim\unit[500]{n\Phi_0/\sqrt{Hz}}$ for
the parameters of Fig.~\ref{fig:F}b.  The results depend on the energy
gap $\Delta_{0,w}$ via
$S_{\Phi,T}^{1/2}\propto{}\sqrt{R_T/\Delta_{0,w}}$ and
$S_{\Phi,\mathrm{pre}}^{1/2}\propto1/\Delta_{0,w}$, so that the
performance is expected to improve with larger-gap superconductors.
Based on these estimates, the performance of the device is in practice
expected to be mostly limited by the external amplifier noise.

\section{Discussion and conclusions}

For weak links short compared to the coherence length, there is little
difference between the normal and superconducting cases, on the level
of the present description. As the junction length increases, the
change in the current-phase relation starts to limit the maximum
achievable modulation of the density of states.  This is reflected in
the decrease of the voltage modulation observed by the tunnel probe.
Moreover, with increasing length of the junction, the gap edge
singularities of the BCS DOS transform to smaller peaks, in a way that
is sensitive to the finite size of the weak link.

The long perturbation decay length at gap edges has an impact on how
generic the results discussed here are in practice, even within the
mean field approximations.  In reality, even if the superconducting
wire is well described by the quasi-1D equations used above, the
boundary conditions at $x=0,L$ may not be as accurate. Namely, the
contact region often has nontrivial 3D structure, and the phase
gradient also extends to the terminals. In such cases, we expect that
in the interior of the wire the results will follow
Eq.~\eqref{eq:bulkg}, but deviations appear at the gap edges, which in
practice are likely to be sensitive to the details not necessarily
described with a single parameter $r$.


In summary, we discussed the current-phase relation, density of
states, and free energy of superconducting wires, focusing on points
relevant for the S-SQUIPT application.  The results point out that
superconducting material is a viable choice, provided the junction
length does not significantly exceed the coherence length, in order
to retain sensitivity and avoid magnetic hysteresis.

We acknowledge funding from the MIUR-FIRB2013 - Project Coca (Grant
No. RBFR1379UX) and the European Research Council under
the European Union’s Seventh Framework Program (FP7/2007- 2013)/ERC
Grant agreement No. 615187-COMANCHE.

\bibliography{squiptdos}

\appendix

\section{Pseudo-arclength continuation}
\label{sec:arclength}

For completeness, we describe here the application of pseudo-arclength
continuation on the self-consistency equation.  The method generates a
set of values $(\Delta_k, \varphi_k)$, tracing a curve of
solutions. The next point $(\Delta_{k+1},\varphi_{k+1})$ is generated from
the previous by solving
\begin{subequations}
  \label{eq:arclength}
  \begin{gather}
  F'[\Delta_{k+1}, \varphi_{k+1}] = 0
  \,,
  \\
  \label{eq:arclength-cons}
  \begin{split}
  s &= (2-\theta)\dot{\varphi}_k(\varphi_{k+1} - \varphi_k) +
  \\
  &+\theta\Re\int_0^L\dd{x} \dot{\Delta}_k^*(x)[\Delta_{k+1}(x) - \Delta_k(x)]
  \,.
  \end{split}
  \end{gather}
\end{subequations}
Here, $F'[\Delta,\varphi]=0$ denotes the set of equations
\eqref{eq:usadel}--\eqref{eq:kl} taking $\Delta(x)$ and the phase
difference $\varphi$ as unknowns.  The value of $\varphi_{k+1}$ is
fixed by the pseudo-arclength constraint~\eqref{eq:arclength-cons}, where the parameter $s>0$ is
an arc-length constant and $\theta\in[0,2]$ a weight factor.  The
tangent approximants can be taken as
$\dot{\varphi}_k=(\varphi_k-\varphi_{k-1})/\delta$,
$\dot{\Delta}_k=(\Delta_k-\Delta_{k-1})/\delta$, with
$\delta^2 = \theta ||\Delta_{k} - \Delta_{k-1}||_2^2 +
(2-\theta)|\varphi_k-\varphi_{k-1}|^2$.
Equations~\eqref{eq:arclength} are of similar complexity as the
self-consistency equation $F'[\Delta]=0$, and can be solved for
$(\Delta_{k+1},\varphi_{k+1})$ using standard nonlinear solvers,
given a spatial discretization of $\Delta$.

\section{Riccati parameterization}

Equations~\eqref{eq:replica-action},\eqref{eq:bc-action} can be written in a Riccati parameterization,
$e^{i\chi}\sin\tilde{\theta}\equiv2\gamma/(1+|\gamma|^2)$:
\begin{align}
  {\cal F}_0
  &=
  |\Delta|^2\ln\frac{T}{T_c}
  +
  2\pi{}T
  \sum_{\omega_n>0}\{
  \frac{|\Delta|^2}{\omega_n}
  +
  \frac{2|\partial_x\gamma|^2}{(1 + |\gamma|^2)^2}
  \\\notag&\quad
  +
  4
  \frac{\omega_n|\gamma|^2 - \Re[\Delta^*\gamma]}{1 + |\gamma|^2}
  \}
  \,,
\end{align}
and
\begin{align}
  F_b
  &=
  \frac{2\pi{}TD}{r}\sum_{\omega_n>0}\sum_{j=\pm}
  \frac{2|\gamma(x_j) - \gamma_{Sj}|^2}{(1 + |\gamma(x_j)|^2)(1 + |\gamma_{Sj}|^2)}
  \,.
\end{align}
This form has some advantages for numerical implementation.
Moreover, the well-known connection to the Ginzburg--Landau functional
\cite{gorkov59b,yurkevich2001-nsm} is straightforward in this form.
The minimum of ${\cal F}_0$ vs.
$\gamma$ is
$\gamma(x,\omega_n)\approx{}\Delta(x)/(2\omega_n)$, up to corrections
$\propto|\partial_x^2\gamma|,|\Delta|^3$. Neglecting the corrections,
substituting the leading term back in, and expanding in small
$|\Delta|$ yields
\begin{align}
  {\cal F}_0
  \approx
  \frac{\pi}{8T}|\partial_x\Delta|^2
  -
  |\Delta|^2\ln\frac{T_c}{T}
  +
  \frac{7\zeta(3)}{16\pi^2T^2}|\Delta|^4
  \,,
\end{align}
the GL functional.

\end{document}